\documentstyle[epsfig,aps]{revtex}

\draft

\draft

\begin{document}

\large

\title{Nonextensive statistics and incomplete information}

\author{Qiuping A. Wang}
\address{Institut Sup\'{e}rieur des Mat\'{e}riaux du Mans, 44, Av.
Bartholdi, 72000 Le Mans, France \\ awang@ismans.univ-lemans.fr}


\maketitle

\begin{abstract}
We comment on some open questions and theoretical peculiarities in
Tsallis nonextensive statistical mechanics. It is shown that the
theoretical basis of the successful Tsallis' generalized
exponential distribution shows some worrying properties with the
conventional normalization and the escort probability. These
theoretical difficulties may be avoided by introducing an so
called incomplete normalization allowing to deduce Tsallis'
generalized distribution in a more convincing and consistent way.
\end{abstract}


\newpage

\section{Introduction}
It is well known that Boltzmann-Gibbs statistics ($BGS$) is
inadequate for treating some complex systems. These are systems
with complex or long term interactions and correlations, systems
showing often distribution laws different from the usual ones
(Gauss, Poisson), systems in chaotic or fractal states and often
related to nonextensive phenomena in energy, entropy, heat, and
other quantities. Some examples of the failures of $BGS$ are given
in reference\cite{Tsal3}. We see that we need a new statistical
mechanics fundamentally different from $BGS$. We can even
conjecture that a new kind of statistical theory may be necessary
for complex random phenomena because the validity of the actual
statistical method is subject to some ideal
conditions\cite{Wang00}. A brief discussion on this topic will be
given in the present paper.

In 1988, in a historical paper\cite{Tsal1,Tsal2}, Tsallis founded
a nonextensive statistical mechanics which, in its most recent
version, gives following canonical distribution functions :
\begin{equation}                                        \label{a1}
p_i=\frac{[1-(1-q)\frac{\beta}{\sum_{i}^w p_i^q}
(e_i-U)]_\dag^\frac{1}{1-q}}{Z}
\end{equation}
with
\begin{equation}                                        \label{a2}
Z=\sum_{i}^w[1-(1-q)\frac{\beta}{\sum_{i}^w p_i^q}
(e_i-U)]_\dag^\frac{1}{1-q}
\end{equation}
where $[y]_\dag=y$ if $y\geq 0$ and $[y]_\dag=0$ otherwise
(Tsallis cut-off). $i$ is a state point in phase space, $w$ the
number of all accessible phase space points for the considered
system, $e_i$ the energy of the system at the state $i$, $\beta$
the Lagrange multiplier of the constraint on the internal energy
$U$ given by
\begin{equation}                                        \label{a2a}
U=\frac{\sum_{i}^w p_i^qe_i}{\sum_{i}^w p_i^q}=\frac{\sum_{i}^w
p_i^qe_i}{Z^{1-q}}
\end{equation}
which is used by Tsallis to overcome some theoretical
difficulties\cite{Tsal2,Penni}. Eqs. (\ref{a1}) and (\ref{a2})
mean
\begin{equation}                                        \label{a3}
\sum_{i}^wp_i=1,
\end{equation}
which is logical if we consider the fact that $w$ represents all
possible states of the system. It should be noticed that, in the
limit $q=1$, Eq. (\ref{a1}) becomes $BGS$, i.e.
$p_i=\frac{1}{Z}exp(-\beta e_i)$.

We refer to the generalized exponential distribution Eq.
(\ref{a1}) as Tsallis distribution function ($TDF$) which is
proved to be indeed very useful and efficacious in describing
some systems with complex or long term interactions. Many
successful and convincing applications was published over the
last 10 years concerning different systems showing peculiar
distribution laws (the reader is referred to references
\cite{Tsal3,Tsal4,Tsal5} for updated comments on this subject).

In spite of the successes of $TDF$, researchers continue to work
on its theoretical foundation, because there are still various
open questions. One notes peculiar theoretical properties which
sometimes are not very easy to understand and deserve to be
investigated. In this paper, we would like to discuss some
fundamental aspects of the last version of Tsallis theory and
present some observations. We also show a possible solution to
the problems with a modifications in the theoretical basis of
nonextensive statistics.

\section{Brief history of $TDF$}
Let us begin by a brief review of the historical path of $TDF$.

In 1988, Tsallis proposed the following generalized statistical
entropy :
\begin{equation}                                        \label{b1}
S=-k\frac{1-\sum_{i=1}^{w}p_i^q}{1-q} , (q \in R)
\end{equation}
which he maximized according to Jaynes principle with two
constraints : Eq. (\ref{a3}) and the mathematically suitable
expectation value of energy U :
\begin{equation}                                        \label{b2}
U=\sum_{i}^wp_ie_i.
\end{equation}
This definition is completely legitimate if $p_i$ is considered
as probability. This approach led to the first version of $TDF$,
i.e.
\begin{equation}                                        \label{b2a}
p_i=\frac{[1-(q-1)\beta e_i]_\dag^\frac{1}{q-1}}{Z}
\end{equation}
with
\begin{equation}                                        \label{b2b}
Z=\sum_{i}^w[1-(q-1)\beta e_i]_\dag^\frac{1}{q-1}
\end{equation}

Later, it was found that $Tsallis-1$ had some shortcomings such as
the thermodynamic instability of entropy and the incapacity to
deduce some power-law distributions
\cite{Tsal2,Tsal6,Rams,Lesc,Guerberoff96}. But quite before the
first published criticisms \cite{Rams,Guerberoff96} on this
theory, another version of $TDF$ was proposed by Curado and
Tsallis \cite{Tsal7}. This second version, say, Curado-Tsallis,
replaced Eq.(\ref{b2}) by :
\begin{equation}                                        \label{b3}
U=\sum_{i}^wp_i^qe_i.
\end{equation}
Curado-Tsallis formalism allows to get a mathematical elegance
with the conventional Legendre transformation of thermodynamics :

\begin{equation}                                        \label{b4}
\frac{\partial{S}}{\partial{U}}=\frac{1}{T}=k\beta
\end{equation}
where $T$ is the absolute temperature. This elegance is lost by
Tsallis-1 version which gives :

\begin{equation}                                        \label{b5}
\frac{\partial{S}}{\partial{U}}=\frac{1}{T}=Z^{1-q}k\beta
\end{equation}
where $U$ is no longer a simple function dependent only on $Z$
just like in the $BGS$ case\footnote{In $Tsallis-1$ version, we
have $U=-\frac{1}{qZ}
\frac{\partial}{\partial{\beta}}\sum_i[1-(q-1)\beta
e_i]_\dag^{\frac{q}{q-1}}$, which nevertheless tends to $BGS$
relation $U=-\frac{\partial}{\partial{\beta}}lnZ$ when
$q\rightarrow 1$.}.

The Curado-Tsallis formalism once again gives Eq.(\ref{b2a}) with
a change $(q-1)\rightarrow (1-q)$, and is successfully applied to
systems showing non gaussian distribution laws as soon as its
proposal. Nevertheless, the anomalous relation between the
normalization Eq.(\ref{a3}) and the averages Eq.(\ref{b3}) remains
an open question. Rigorously speaking, supposing Eq.(\ref{a3}), we
logically write Eq.(\ref{b2}) for an observable average. In other
words, Eq.(\ref{b3}) means that $p_i^q$ is the observable
probability but is not normalized and that $p_i$ is an imaginary
probability which is normalized but never used in practice, i.e.
$p_i$ is not observable if Eq.(\ref{a3}) holds.

This second formalism, according to some
authors\cite{Tsal2,Penni,Guerberoff96}, shows some fundamental
difficulties. For examples : 1) the average of a constant is not
constant; 2) the total energy of two systems without any
interaction is not the sum of the energies of each system; 3) the
zeroth law does not hold; 4)the invariance of probability with
uniform energy translation is missing.

The third or last version of $TDF$ is that mentioned at the
beginning of this paper. It replaces the unnormalized average by a
normalized one Eq.(\ref{a2a}) and so resolved the above mentioned
problems.

When writing this manuscript, we saw another proposal by Martinez
et al \cite{Marti} concerning the energy constraint for entropy
maximization. They propose using
\begin{equation}                                        \label{b3a}
\sum_{i}^wp_i^q(e_i-U),
\end{equation}
as the constraint but with $U$ given by Eq.(\ref{a2a}). This
approach overcomes a mathematical difficulty of the third version
of $TDF$, that is, the auxiliary function $A(p)$ for entropy
maximization :
\begin{equation}                                        \label{b3b}
A=-\frac{1-\sum_{i=1}^{w}p_i^q}{1-q}-\alpha(\sum_{i}^wp_i-1)-
\beta (\frac{\sum_{i}^w p_i^qe_i}{\sum_{i}^w p_i^q}-U)
\end{equation}
does not necessarily have a maximum. Because, first,
$\frac{d^2A}{dp_idp_j}$ is not diagonal, i.e.,
$\frac{d^2A}{dp_idp_j}\neq 0$ when $i\neq j$. Second, even if we
calculate only $\frac{d^2A}{dp_i^2}$, we get :
\begin{equation}                                        \label{b3c}
\frac{d^2A}{dp_i^2}=-qp_i[Z^{1-q}- q\beta
p_i^{2q-1}Z^{2q-2}(e_i-U)].
\end{equation}
To ensure $\frac{d^2A}{dp_i^2}<0$ for a maximum entropy, we have
to put $[Z^{1-q}- q\beta p_i^{2q-1}Z^{2q-2}(e_i-U)]>0$, which,
evidently, can not be always valid. Martinez et al overcame this
difficulty with Eq.(\ref{b3a}) leading to
$\frac{d^2A}{dp_idp_j}=0$ for $i\neq j$ and
$\frac{d^2A}{dp_i^2}=-qp_iZ^{1-q}<0$. The entropy maximum is
ensured.

This maximization still leads to Eq.(\ref{a1}) as probability
distribution. In fact, this proposal is a composition of the
Curado-Tsallis maximization with $\sum_{i}^wp_i^qe_i$ and another
constraint $\sum_{i}^wp_i^qU$. The latter can be called invariance
constraint since its only role is to ensure the invariance of the
resulted distribution with respect to uniform translation of the
energy levels $e_i$. Due to the same distribution functions, this
scenario has the same essential characteristics as the last
version of $TDF$, about which we will discuss some open questions
in the following section.

\section{Some questions about $TDF$} \label{prob}

\begin{enumerate}
\item
Let us begin by presenting an observation about the maximum
entropy principle of Jaynes. This principle asserts that, in
order to obtain the correct probability distribution, it suffices
to introduce into the entropy maximization physical conditions or
constraints related to observable quantities, that is, the
normalization of the probability, the expectation, variance or
higher moments. Supposed Eq.(\ref{a3}) as normalization, the only
observable expectation value or higher moments must be defined
with Eq.(\ref{b2}). The introduction of Eq.(\ref{b3}) or
Eq.(\ref{a2a}) does not conforms with this principle, because we
do not know whether these averages really represent observable
quantities or not. One may say that the successful applications
of these formalisms confirm the observability of these averages.
But in this case, $p_i$ becomes non observable in turn. So the
normalization Eq.(\ref{a3}) should disappear in the entropy
maximization. This observability question deserves to be
carefully studied. Indeed, if necessary, we could reject any old
principle and introduce new ones under the condition that the
resulted theory be useful and self-consistent.

\item
The second question was discussed by Raggio \cite{Raggio}. If the
expectation Eq. (\ref{a2a}) satisfies the constraint associated
with the linearity in the observables, i.e.
\begin{equation}                                        \label{b6}
\overline{x+y}=\overline{x}+\overline{y}
\end{equation}
for two independent observables \^x and \^y, it does not satisfy
the linearity in the state (or distribution), i.e.
\begin{equation}                                        \label{b7}
\overline{x}[\lambda p(1)+(1-\lambda)p(2)]= \lambda
\overline{x}[p(1)]+(1-\lambda)\overline{x}[p(2)]
\end{equation}
where $p(1)$ and $p(2)$ are two normalized probability
distributions, nor does Eq. (\ref{b3}) which violates even
Eq.(\ref{b6}). The origin of these violations is logically the
{\it discrepancy between the average Eq. (\ref{a2a}) (or
(\ref{b3})) and the normalization Eq. (\ref{a3}) } according to
which the joint probability is $p(x+y)=p(x)p(y)$ and the
probability summation is $p(1+2)=p(1)+p(2)$. If the average is
calculated with $p^q$ instead of $p$, the problem is evident
since $p^q$ is not normalized and $p^q(1+2) \neq p^q(1)+p^q(2)$.
Indeed,
\begin{eqnarray}                                        \label{b7a}
\overline{x}[\lambda p(1)+(1-\lambda)p(2)] &=&\frac{\sum_i[\lambda
p_i(1)+(1-\lambda)p_i(2)]^qx_i} {\sum_i[\lambda
p_i(1)+(1-\lambda)p_i(2)]^q}  \\   \nonumber
&\neq&\lambda\frac{\sum_ip_i^q(1)x_i}{\sum_ip_i^q(1)}
+(1-\lambda)\frac{\sum_ip_i^q(2)x_i}{\sum_ip_i^q(2)}.
\end{eqnarray}
Eqs. (\ref{b6}) and (\ref{b7}) are to be satisfied for that \^x,
\^y, $p(1)$ and $p(2)$ be observable and have physical
signification. This problem naturally leads us to the following
question : does the observability of the normalization Eq.
(\ref{a3}) is incompatible with that of the expectation value
given by Eq. (\ref{a2a}) or (\ref{b3})?

\item
The third question we would like to discuss is about the absence
of analytic energy correlation. We know that the correlations in
entropy can be calculated {\it a priori} with mathematical rigor.
But the correlation in any other observable quantities can not be
calculated with the average Eq. (\ref{a2a}).

Let us suppose an isolated system $C$ composed of two subsystems
$A$ and $B$ of which the distributions satisfy
\begin{eqnarray}                                    \label{b8}
p_{ij}(C)=p_i(A)p_j(B).
\end{eqnarray}
In nonextensive statistics, this hypothesis of multiplication law
means that $A$ and $B$ are correlated and gives the correlation
term associated with energy. From Eq. (\ref{a1}), (\ref{a2}) and
(\ref{a2a}), we straightforwardly obtain :
\begin{eqnarray}                                    \label{b9}
e_{ij}(A+B)-U(A+B) & = &
[e_i(A)-U(A)]\sum_ip_{i}^q(B)+ [e_j(B)-U(B)]\sum_jp_j^q(A) \\
\nonumber
 & + & (q-1)\beta[e_i(A)-U(A)][e_j(B)-U(B)].
\end{eqnarray}
Without additional hypothesis, this equality does not lead to any
explicit relation between the hamiltonians $H(A+B)$ and $H(A)$ or
$H(B)$ neither for the micro-state values $e_i$ nor for the
average $U$. Tsallis and coworkers \cite{Tsal2,Abe99} proposed
neglecting the correlation term of any observable between the
subsystems. So one can write :

\begin{eqnarray}                                    \label{b10}
e_{ij}(A+B) & = & e_i(A) + e_j(B)
\end{eqnarray}
and
\begin{eqnarray}                                    \label{b11}
U(A+B) & = & U(A) + U(B).
\end{eqnarray}
On the basis of this {\it extensive energy approximation}, the
zeroth law of thermodynamics is claimed to be established for the
foundation of nonextensive thermodynamics\cite{Abe99,Mart01}. This
problem will be discussed below

Our question about Eqs. (\ref{b10}) and (\ref{b11}) is : if in
general the correlation (nonextensive) terms of whatever
observable or interactions can be neglected, what is the origin
of the nonextensivity of entropy? We know that entropy should be
a continuous function of the distributions which in turn are
continuous functions of the observables.

In addition, allowing Eqs.(\ref{b10}) and (\ref{b11}), we lose
the equalities Eqs.(\ref{b9}) and Eq.(\ref{b8}), which are
crucial for the nonextensive theory. If Eq.(\ref{b8}) fails, we
cannot in fact find even the entropy correlation given by
\begin{equation}                                \label{b12}
S(A+B)=S(A)+S(B)+\frac{1-q}{k}S(A)S(B).
\end{equation}

\item
Zeroth law of thermodynamics can not be established without
neglecting the correlation energy. That is to say that the
Lagrange multiplier $\beta$ is no longer a "meter" measuring the
thermodynamic equilibrium. We will show why.

We take once again the above mentioned isolated system $C$
composed of two subsystems $A$ and $B$ in equilibrium. From Eq.
(\ref{b12}), we get, for a small variation of the total entropy :
\begin{eqnarray}                                    \label{b13}
\delta S(A+B) & = & [1+\frac{1-q}{k}S(B)]\delta S(A)+
[1+\frac{1-q}{k}S(A)]\delta S(B) \\ \nonumber
    & = & [1+\frac{1-q}{k}S(B)]
\frac{\partial S(A)}{\partial U(A)} \delta U(A) +
[1+\frac{1-q}{k}S(A)] \frac{\partial S(B)}{\partial U(B)} \delta
U(B).
\end{eqnarray}
Because $\delta S(A+B)=0$, we get
\begin{eqnarray}                                    \label{b14}
[1+\frac{1-q}{k}S(B)] \frac{\partial S(A)}{\partial U(A)} \delta
U(A) + [1+\frac{1-q}{k}S(A)] \frac{\partial S(B)}{\partial U(B)}
\delta U(B)=0
\end{eqnarray}
Now, in order to proceed, we need the relation between $U(A+B)$
and $U(A)$ or $U(B)$ but it does not exist. As mentioned above,
$U(A+B)$ can not be expressed in $U(A)$ and $U(B)$. So no
relation can be found between $\delta U(A)$ and $\delta U(B)$ and,
as a consequence, the derivative $\frac{\partial S}{\partial U}$
can not be calculated from $\delta S(A+B)=0$. So supposing
$U(A+B)= U(A) + U(B)$ or $\delta U(A)=-\delta U(B)$ proposed by
Abe and Martinez et al \cite{Abe99,Mart01}, we obtain
\begin{eqnarray}                                    \label{b15}
[1+\frac{1-q}{k}S(B)] \frac{\partial S(A)}{\partial U(A)} =
[1+\frac{1-q}{k}S(A)] \frac{\partial S(B)}{\partial U(B)}
\end{eqnarray}
or
\begin{eqnarray}                                    \label{b16}
Z^{q-1}(A)\beta(A) = Z^{q-1}(B)\beta(B)
\end{eqnarray}
which is the generalized zeroth law with $Z^{q-1}\beta$ instead
of $\beta=1/kT$ as the measure of the equilibrium. Due to this
approximate zeroth law, the implicit distribution function Eq.
(\ref{a1}) becomes explicit for systems in thermal equilibrium
since it can be recast as
\begin{equation}                                        \label{b17}
p_i=\frac{[1-(1-q)\lambda(e_i-U)]_\dag^\frac{1}{1-q}}{Z}
\end{equation}

where $\lambda=\beta Z^{q-1}$ is now an independent thermodynamic
variable. So the theory is reconciled with the old notion of
thermal equilibrium to the detriment of the correlation between
the components of the system. In our opinion, two important things
are lost in this treatment : a) the generality of the
nonextensive theory which should be able to tackle correlated
problems in taking into account the interactions; b) the
generality of the zeroth law which should hold within a theory
without any condition, or all thermodynamic laws will become
approximate. As a matter of fact, as mentioned above and argued
by Guerberoff {\it et al}\cite{Guerberoff96}, with additive
energy or non interacting subsystems, Eq.(\ref{b8}) and
Eq.(\ref{b12}) do not hold so that the zeroth law Eq.(\ref{b16})
can not be established. Very recently, a new point of view about
thermal equilibrium and nonextensivity shows that Eq.(\ref{b8})
or Eq.(\ref{b12}) is required by the existence of thermal
equilibrium with Tsallis entropy and should be regarded as a
basic assumption of the statistics for equilibrium nonextensive
systems. So that appropriate energy nonextensivity satisfying
Eq.(\ref{b8}) is absolutely necessary for the validity of zeroth
law within Tsallis statistics\cite{Abe01a,Wang02}. We will come
back to this issue later in this paper.

\item
Now we discuss a mathematical problem. From Eq. (\ref{a1}),
we can write :
\begin{eqnarray}                                        \label{b18}
Z&=&\sum_{i}^w[1-(1-q)\lambda (e_i-U)]_\dag^\frac{1}{1-q}
\\ \nonumber &=&\sum_{i}^w[1-(1-q)\lambda
(e_i-U)]_\dag^{\frac{q}{1-q}+1} \\ \nonumber
&=&Z^q\sum_{i}^wp_i^q[1-(1-q)\lambda (e_i-U)]_\dag
\end{eqnarray}
Considering the average defined in Eq. (\ref{a2a}), Eq.
(\ref{b18}) can be recast as :
\begin{eqnarray}                                        \label{b19}
Z&=&\sum_{i}^w[1-(1-q)\lambda (e_i-U)]_\dag^\frac{q}{1-q}
\end{eqnarray}
or
\begin{eqnarray}                                        \label{b20}
\sum_i^w[1-(1-q)\lambda(e_i-U)]_\dag^\frac{1}{1-q}=
\sum_i^w[1-(1-q)\lambda(e_i-U)]_\dag^\frac{q}{1-q}.
\end{eqnarray}
This equality is a basic relation of the theory and must hold for
arbitrary value of $q$, $w$, $\beta$ and $e_i$. Now let us apply
it to calculate the inverse temperature $\beta$.

From Eq. (\ref{b19}), we can write
\begin{eqnarray}                                        \label{b21}
\sum_{i}^w p_i^q=Z^{1-q}
\end{eqnarray}
and
\begin{equation}                                        \label{b22}
S=k\frac{Z^{1-q}(U)-1}{1-q},
\end{equation}
Then we calculate the following derivative :
\begin{eqnarray}                                        \label{b23}
\frac{dS}{dU}&=&\frac{k}{Z^q}\frac{dZ}{dU}.
\end{eqnarray}
First we take the $Z$ given by Eq. (\ref{a2}), we obtain :

\begin{eqnarray}                                        \label{b24}
\frac{dS}{dU} = k\lambda Z^{1-q}=k\beta.
\end{eqnarray}
But if we take the $Z$ of Eq. (\ref{b19}), we obtain
\begin{eqnarray}                                        \label{b25}
\frac{dS}{dU}=\frac{qk\beta}{Z}\sum_{i}^w[1-(1-q)\lambda
(e_i-U)]_\dag^\frac{2q-1}{1-q}
\end{eqnarray}
which means
\begin{eqnarray}                                        \label{b26}
Z=q\sum_{i}^w[1-(1-q)\lambda (e_i-U)]_\dag^\frac{2q-1}{1-q}.
\end{eqnarray}
If we put Eq. (\ref{b26}) into Eq. (\ref{b23}) and continue in
this way for $n$ times, we will find
\begin{eqnarray}                                        \label{b27}
Z&=&\sum_{i}^w[1-(1-q)\lambda(e_i-U)]_\dag^\frac{q}{1-q} \\
\nonumber
 &=& q\sum_{i}^w[1-(1-q)\lambda(e_i-U)]_\dag^\frac{2q-1}{1-q}\\
\nonumber
&=& q(2q-1)\sum_{i}^w[1-(1-q)\lambda(e_i-U)]_\dag^\frac{3q-2}{1-q}\\
\nonumber
&=& q(2q-1)(3q-2)\sum_i^w[1-(1-q)\lambda(e_i-U)]_\dag^\frac{4q-3}{1-q}\\
\nonumber &=& q(2q-1)(3q-2)...(nq-n+1)
\sum_i^w[1-(1-q)\lambda(e_i-U)]_\dag^\frac{(n+1)q-n}{1-q}
\end{eqnarray}
with $n=0, 1, 2 ... $. We create in this way a series of
equalities which seem not to hold. For example, if we take the
second equality of Eq. (\ref{b27}) and let $q\rightarrow 0$, the
right-hand side will tend to zero and the left-hand side to
$\sum_i^w1=w$. The result is $w\rightarrow 0$. This same result
can also be obtained for $q\rightarrow \frac{1}{2}$ if we take the
third equality of Eq.(\ref{b27}) and for $q\rightarrow
\frac{2}{3}$ with the forth equality and so on. These singular
points in $q$ value do not conform with the hypothesis that Eq.
(\ref{b20}) is a general relation of the theory. It seems to us
that these equalities are valid only when $q\rightarrow 1$ and
$Z$ becomes the $BGS$ partition function.

We can also study the equality Eq.(\ref{b20}) in another way. Let
us suppose that \^{x} is a continuous variable within $0<x<
\infty$. So $Z$ sometimes can be given by
\begin{eqnarray}                                        \label{b28}
Z=\int_0^\infty [1-(1-q)\lambda(x-U)]_\dag^\frac{q}{1-q}dx.
\end{eqnarray}
or
\begin{eqnarray}                                        \label{b29}
Z=\int_0^\infty [1-(1-q)\lambda(x-U)]_\dag^\frac{1}{1-q}dx.
\end{eqnarray}
In this case, we should put $q>1$ for $Z$ to be calculated when
$x$ is large. The integration of Eq.(\ref{b28}) is always finite.
But Eq.(\ref{b29}) needs $q<2$ to be finite. If $q>2$, the $Z$ of
Eq.(\ref{b28}) can be calculated while that of Eq.(\ref{b29})
diverges. This paradox naturally disappears for $q\rightarrow 1$.

\item
Another problem concerning the relation between the average $U$
and the micro-state value $e_i$ or $Z$ arises due to
Eq.(\ref{b20}). Usually, the $U-Z$ relation
($U=-\frac{\partial}{\partial\beta}lnZ$ in $BGS$) can be found by
introducing the distribution function Eq.(\ref{a1}) into the
average calculus Eq.(\ref{a2a}):
\begin{eqnarray}                                       \label{b29a}
U&=&\frac{\sum_{i} p_i^qe_i}{Z^{1-q}} \\ \nonumber
&=&\frac{1}{Z}\sum_{i}e_i[1-(1-q)\lambda
(e_i-U)]_\dag^\frac{q}{1-q}
\\ \nonumber &=&-\frac{1}{Z} \{ \frac{\partial}{\partial\lambda}
\sum_{i}[1-(1-q)\lambda (e_i-U)]_\dag^\frac{1}{1-q}
-\sum_{i}U[1-(1-q)\lambda (e_i-U)]_\dag^\frac{q}{1-q} \}
\\   \nonumber
&=&-\frac{1}{Z} \{\frac{\partial Z}{\partial\lambda}-UZ  \} \\
\nonumber &=& -\frac{1}{Z}\frac{\partial Z}{\partial\lambda}+U
\end{eqnarray}
which leads to, instead of the expected $U-Z$ relation,
\begin{eqnarray}                                      \label{b29a1}
\frac{\partial Z}{\partial\lambda} = 0.
\end{eqnarray}
as well as
\begin{eqnarray}                                      \label{b29b}
\frac{\partial S}{\partial\lambda}
=\frac{\partial}{\partial\lambda}k\frac{Z^{1-q}-1}{1-q}=
k\frac{1}{Z^q}\frac{\partial Z}{\partial\lambda}= 0.
\end{eqnarray}
So the micro-macro relation is impossible to be found if no
mechanical quantity can be calculated from its microstate values.

In addition, Eq.(\ref{b29a1}) gives rise to another problem
similar to that discussed in the precedent part. Eq.(\ref{b29a1})
can be easily verified if we take the standard $Z$ given by
Eq.(\ref{a2}). But If we take the $Z$ in Eq.(\ref{b19}), we get
the following relation

\begin{eqnarray}                                        \label{b29c1}
\sum_{i}^w(e_i-U)[1-(1-q)\lambda (e_i-U)]_\dag^\frac{2q-1}{1-q}=0
\end{eqnarray}
or
\begin{eqnarray}                                       \label{b29c2}
U=\frac{\sum_{i}^we_i[1-(1-q)\lambda
(e_i-U)]_\dag^\frac{2q-1}{1-q}}{\sum_{i}^w[1-(1-q)\lambda
(e_i-U)]_\dag^\frac{2q-1}{1-q}}=
\frac{\sum_{i}^we_ip_i^{2q-1}}{\sum_{i}^wp_i^{2q-1}}.
\end{eqnarray}
If we repeat the same reasoning with the $Z$ of Eq.(\ref{b26}), we
get
\begin{eqnarray}                                       \label{b29c3}
U=\frac{\sum_{i}^we_ip_i^{3q-2}}{\sum_{i}^wp_i^{3q-2}}.
\end{eqnarray}
We can continue in this way with Eq.(\ref{b29a}) and obtain :
\begin{eqnarray}                                       \label{b29c4}
U& = &\frac{\sum_{i}^we_ip_i^q}{\sum_{i}^wp_i^q}
\\ \nonumber
&=&\frac{\sum_{i}^we_ip_i^{2q-1}}{\sum_{i}^wp_i^{2q-1}} \\
\nonumber &=&\frac{\sum_{i}^we_ip_i^{3q-2}}{\sum_{i}^wp_i^{3q-2}} \\
\nonumber &=&\frac{\sum_{i}^we_ip_i^{4q-3}}{\sum_{i}^wp_i^{4q-3}}
\\ \nonumber
...&=& \frac{\sum_{i}^we_ip_i^{nq-n+1}}{\sum_{i}^wp_i^{nq-n+1}}.
\end{eqnarray}

which means $\sum_{i}^w(e_i-U)=0$ or $U=\sum_{i}^we_i/w$ if
$q=\frac{n-1}{n}$ with $n=1,2,3 ...$. i.e., we are led to the
microcanonical case. On the other hand, if we calculate $U$ from
its definition Eq.(\ref{a2a}) for $q=1/2$, we will be led to
\begin{eqnarray}                                       \label{b29d}
U=\frac{1}{Z}(\sum_{i}^we_i-\frac{\lambda}{2} \sum_{i}^we_i^2 +
\frac{\lambda}{2}U\sum_{i}^we_i)
\end{eqnarray}
or
\begin{eqnarray}                                       \label{b29e}
U=(\sum_{i}^we_i-\frac{\lambda}{2}\sum_{i}^we_i^2)
/(Z-\frac{\lambda \sum_{i}^we_i}{2}),
\end{eqnarray}
which does not seem to be a microcanonical case.

These mathematical problems discussed above seem to be related
directly to the fact that the $TDF$ Eq.(\ref{a1}) depends only on
the difference $e_i-U$. So Eq.(\ref{a1}) can be regarded as the
relative probability distribution with respect to the average
energy of the system, but not the real probability with respect
to a suitable zero-energy we must choose due to the energy
translation variance of $TDF$. We will come back to this problem
at the end of the present paper. We also wonder if the
$q$-independent results for generalized ideal gas\cite{Marti}
have something to do with this relative probability combined with
extensive energy approximation for ``non interacting particles".

\end{enumerate}

We have discussed some questionable points in the last version of
$TDF$. Some other discussions concerning the relation between
$TDF$ of the third version and the law of large number are given
in reference \cite{Lacour}. Of course, these are questions that we
have to study carefully. For the moment, we do not see how to get
out of these theoretical difficulties if we stay in the formalism
with the conventional normalization and the expectation value
Eq.(\ref{a2a}). In what follows, we will show a possible way out.
The main idea is to introduce into Tsallis theory the hypothesis
of incompleteness of our knowledge (information) about
nonextensive systems.

\section{Some considerations concerning statistics}

Very recently, a possible alternative theoretical basis for $TDF$
was proposed \cite{Wang00,Wang01}. The new formalism is based on
a reflection about the conditions of physical application of the
standard probability theory which is sometimes referred to as
Kolmogorov probability theory \cite{Reny66}. This reflection leads
to a mathematically simpler and coherent nonextensive framework
being capable of avoiding the problems discussed above.

This new theoretical framework is referred to as {\it incomplete
statistics (IS)}. The basic idea of $IS$ is that {\it our
knowledge about the states (their position and total number in
the phase space) of a system may be incomplete and non exact.}
This is true at least for complex systems with unknown space-time
correlations which can not be exactly described with analytic
methods. So the equation of motion must be incomplete in the
sense that some interactions are missing and its solution can not
yield complete knowledge about the systems. On the other hand,
Kolmogorov probability theory is founded on the hypothesis that
we know all the possible states of the studied system or that the
maximal information is complete and can provide definite answers
for all questions that can be asked of the system. When we carry
out a summation or an integration of probability in phase-space,
we suppose that this is done over the possible states which can be
determined by the equation of motion. This assumption is logical
if and only if all interactions and space-time correlations are
well-known or their unknown parts are negligible, as in the case
of $BGS$ or of other successful probabilistic sciences.

\section{Incomplete normalization}

If the incompleteness of our knowledge is not negligible, it is no
longer sure that we sum over the possible states of the system
simply because we do not know all of them. What we can do is to
take the known states or events suggested by the equation of
motion or by our knowledge. Their number, say $v$, may be greater
or smaller than $w$, the real number of all possible states. As a
consequence, the normalization condition is reduced to
\begin{equation}                            \label{c1}
\sum_{i=1}^{v}p_i=Q\neq 1
\end{equation}
where $p_i$ is the true probability which can not sum to one. The
necessity to introduce this nonextensive or nonadditive
probability is first noted by economists\cite{Tsal3}. This
probability in Eq.(\ref{c1}) was referred to by R\'enyi as {\em
incomplete probability distribution}\cite{Reny66} because the
values of the random variable of this distribution do not
constitute a complete (exhaustive) ensemble (i.e. $v<w$).

It should be noticed that $Q$ is a constant depending on the
studied system. Now in order to apply conventional probability
theory, Eq.(\ref{c1}) has to be renormalized to get a calculable
``probability" related to $p_i$ as well as to $Q$ representing
the nature of the system.

Our proposal\cite{Wang00} is to write
\begin{equation}                                 \label{c2}
\sum_{i=1}^{v}\frac{p_i}{Q}=\sum_{i=1}^{v}p_i^q.
\end{equation}
which means
\begin{equation}                                        \label{c3}
\sum_{i=1}^{v}p_i^q=1, (q\in[0,\infty])
\end{equation}
and, for internal energy,
\begin{equation}                                        \label{c4}
U=\sum_{i=1}^{v}p_i^qe_i.
\end{equation}
Since $p_i$ is the true probability and satisfies $0\leq p_i<1$,
we have to set $0<q<\infty$. $q=0$ should be avoided because it
leads to $p_i=0$ for all states. We see from Eq.(\ref{c2}) that
$Q=1$, $Q<1$ and $Q>1$ means $q=1$, $q<1$ and $q>1$,
respectively. $q$ is directly related to $Q$ (e.g., for
microcanonical distribution, $q=\frac{\ln v}{\ln v-\ln Q}$) and,
in this way, to the unknown correlations or information. This may
help to understand empirical $q$ values for nonextensive systems.

\section{Nonextensive statistics}
In this section, we present a method based on the incomplete
normalization Eq.(\ref{c3}) in order to get $TDF$.

\subsection{Nonextensive information and entropy}
On the basis of the hypothesis of the nonextensivity
(nonadditivity) of entropy or other quantities, we proposed using
a generalized logarithm function as a generalized Hartley formula
for the information measure I(N) required to specify one element
in a system containing $N$ elements :
\begin{equation}
I(N)=\frac{N^g-1}{g}                             \label{c5}
\end{equation}
where $g$ is a real number. This means the following
nonadditivity :
\begin{equation}                                       \label{c6}
I(N_1\times N_2)=I(N_1)+I(N_2)+gI(N_1)\times I(N_2).
\end{equation}
where $I(N_1\times N_2)$ is the information needed to specify
simultaneously 2 elements, each being in a subsystem 1 or 2. We
see that the parameter $g$ is a measure of nonextensivity. If
$g\rightarrow 0$, $I(N)\rightarrow lnN$ and the information
becomes extensive (additive). Eq.(\ref{c4}), Eq.(\ref{c5}) or
Eq.(\ref{c6}), plus the other axioms used by
Shannon\cite{Wang00,Shan}, lead to a nonextensive entropy
\begin{equation}                                       \label{c7}
S=-k\sum_{i=1}^{v}p_i^q\frac{p_i^g-1}{g}
=-k\frac{\sum_{i=1}^{v}p_i^{q+g}-1}{g}.
\end{equation}

We should ask that the above entropy recover the Gibbs-Shannon
one $S=-k\sum_{i=1}^{w}p_ilnp_i$ for $q=1$. This constraint on
nonextensive entropy is logical because $q=1$ or $Q=1$ implies a
complete knowledge about the studied system and short range
interactions or correlations. In this case, we do not have any
reason for holding the nonextensivity. So $g=0$ when $q=1$ and
$g$ should be monotonic function of $q$, which ensures the
monotonic $q$-dependence of entropy.

\subsection{generalized distribution function}
For maximum entropy, we write the following auxiliary function :
\begin{equation}                                      \label{c8}
A=\frac{1-\sum_{i=1}^{v}p_i^{q+g(q)}}{g(q)}+\alpha\gamma
\sum_{i}^vp_i^q-\alpha\beta'{\sum_{i}^v p_i^qe_i}.
\end{equation}
Let $\frac{dA}{dp_i}=0$, we obtain
\begin{equation}                                      \label{c9}
p_i=\frac{[1-\frac{\beta'}{\gamma} e_i]_\dag^\frac{1}{g(q)}}{Z}
\end{equation}
with
\begin{equation}                                     \label{c10}
Z^q=\sum_{i}^v[1-\frac{\beta'}{\gamma} e_i]_\dag^\frac{q}{g(q)}.
\end{equation}
Now it should be asked that the distribution Eq.(\ref{c9}) become
the $BGS$ exponential distribution for $q=1$ or $g(q)=0$, which
means that Eq.(\ref{c9}) should be a generalized exponential
function, that is
\begin{equation}                                     \label{c11}
\frac{Z^gp_i^g(e_i)-1}{g}=-\beta' e_i.
\end{equation}
This straightforwardly leads to $\gamma = 1/g(q)$ and
\begin{equation}                                     \label{c12}
p_i=\frac{[1-g(q)\beta' e_i]_\dag^\frac{1}{g(q)}}{Z}
\end{equation}
From Eq.(\ref{c12}), it can be shown that
\begin{equation}                                     \label{c13}
\frac{d^2A}{dp_i^2}=-[g(q)+q]p_i^{g(q)+q-2}.
\end{equation}
If we want that the distribution Eq.(\ref{c9}) be a maximum
entropy (or minimum energy\cite{Mende}) distribution, we should
put
\begin{equation}                                    \label{c14}
g(q)+q>0
\end{equation}
which ensures $\frac{d^2A}{dp_i^2}<0$ for a maximum entropy with
Eq.(\ref{c8}). This means that the curve of $g(q)$ is situated
above the straight line of $g(q)=-q$. If we impose the
monotonicity of $S$ and $g(q)$ with respect to $q$, we also have
\begin{equation}                                    \label{c15}
\frac{dg(q)}{dq}<0.
\end{equation}

Eq.(\ref{c12}) is the generalized nonextensive distribution
function. Comparing it to Eq.(\ref{a1}), we see that $TDF$
corresponds to $g(q)=1-q$. This choice is the simplest one that
satisfies the two conditions Eqs.(\ref{c14}) and (\ref{c15}) and
also yields the simplest nonextensive entropy :
\begin{equation}                                    \label{c16}
S=-k\sum_{i=1}^{v}p_i^q\frac{p_i^{1-q}-1}{1-q}
=-k\sum_{i=1}^{v}\frac{p_i-p_i^q}{1-q}
=k\frac{1-\sum_{i=1}^{v}p_i}{1-q}.
\end{equation}
with the simplest generalized distribution function
\begin{equation}                                    \label{c17}
p_i=\frac{[1-(1-q)\beta' e_i]_\dag^\frac{1}{1-q}}{Z}.
\end{equation}
which is the $TDF$ in $IS$ version. It should be remembered that a
different forms of $g(q)$ may lead to different distributions.

Note that the distribution Eq.(\ref{c17}) is in the same form as
the $TDF$ in Curado-Tsallis version but, due to the incomplete
normalization, has a different partition function given by
\begin{equation}                                    \label{c18}
Z^q=\sum_{i}^v[1-(1-q)\beta' e_i]_\dag^\frac{q}{1-q}.
\end{equation}
The consequences of this change will be discussed in the
following sections.

We would like to mentioned here that
Kaniadakis\cite{Kania1,Kania2} proposed a new generalization of
$BGS$ ($\kappa$-deformed statistics) on the basis of nonlinear
kinetics in low density gas systems with normalized distribution
function $f$. It is argued that the $q$-variance of the
$q$-exponential function of $TDF$ suggests to write $f=p_i^q$ and
so Eqs.(\ref{c3}) and (\ref{c4}). This work gives Eq.(\ref{c17})
as single particle distribution from the nonlinear kinetics
theory. This result is in accordance with the conclusion of
reference \cite{Wang02} that $TDF$ is an exact distribution for
both many-body system (nonextensive) and correlated single body
according to Eq.(\ref{b8}) prescribed by thermodynamic
equilibrium\cite{Abe01a,Wang02}. In this sense, the hypothesis of
{\it low density gas} is no longer necessary, and the
$\kappa$-deformed statistics can be regarded as a valid theory
for any nonextensive gas system in equilibrium having Tsallis
entropy.

\subsection{Generalized distribution and nonextensivity}
We consider again the total system $C$ composed of two subsystems
$A$ et $B$ in interaction. By $p_{ij}(C)$ we denote the
probability that $C$ is at the product state $ij$ while $A$ is at
the state $i$ with a probability $p_i(A)$ and $B$ at $j$ with a
conditional probability $p_{ij}(B\mid A)$. We can write

\begin{eqnarray}                                    \label{d1}
p_{ij}(C)=p_i(A)p_{ij}(B\mid A),
\end{eqnarray}
or
\begin{eqnarray}                                    \label{d2}
p_{ij}^q(C)=p_i^q(A)p_{ij}^q(B\mid A),
\end{eqnarray}
which means
\begin{equation}                                    \label{d3}
\frac{[1-g(q)\beta' e_{ij}(C)]_\dag^\frac{q}{g(q)}}{Z^q(C)}
=\frac{[1-g(q)\beta' e_{i}(A)]_\dag^\frac{q}{g(q)}}{Z^q(A)}
\frac{[1-g(q)\beta' e_{ij}(B)]_\dag^\frac{q}{g(q)}}{Z^q(B\mid A)}
\end{equation}
or
\begin{equation}                                    \label{d4}
e_{ij}(C)=e_{i}(A)+e_{ij}(B)-g(q)\beta' e_{i}(A)e_{ij}(B).
\end{equation}
and, from Eq.(\ref{c4}),
\begin{equation}                                    \label{d5}
U(C)=U(A)+U(B)-g(q)\beta' U(A)U(B).
\end{equation}
Since $S$ is an observable as the others, Eq.(\ref{d5}) must hold
for $S$ as well. Indeed, if we put the multiplication law
Eq.(\ref{d1}) into the entropy Eq.(\ref{c7}), we get
\begin{eqnarray}                                    \label{d6}
S(C)&=&-k\frac{\sum_{ij}p_{ij}^{q+g(q)}(C)-1}{g(q)} \\ \nonumber
&=&-k\frac{\sum_{ij}p_{i}^{q+g(q)}(A)p_{ij}^{q+g(q)}(B\mid
A)-1}{g(q)} \\ \nonumber
&=&-k\frac{\sum_{i}p_{i}^{q+g(q)}(A)-1}{g(q)}
-k\frac{\sum_{j}p_{ij}^{q+g(q)}(B\mid A)-1}{g(q)} \\ \nonumber
&-&\frac{g(q)}{k}k^2\frac{\sum_{i}p_{i}^{q+g(q)}(A)-1}{g(q)}
\frac{\sum_{j}p_{ij}^{q+g(q)}(B\mid A)-1}{g(q)} \\ \nonumber
&=&S(A)+S_i(B\mid A)-\frac{g(q)}{k}S(A)S_i(B\mid A).
\end{eqnarray}
with
\begin{equation}                                     \label{d8}
S_i(B\mid A)=-k\frac{\sum_jp_{ij}^{q+g(q)}(B\mid A)-1}{g}.
\end{equation}
Following the idea of Abe et al\cite{Abe00}, we can define a
conditional entropy of $B$ as follows
\begin{equation}                                    \label{d7}
S(B\mid A)=\sum_i p_i^q(A)S_i(B\mid A).
\end{equation}
From Eq.(\ref{d6}), we simply obtain
\begin{eqnarray}                                    \label{d9}
S(B\mid A)&=&\sum_i p_i^q(A)\frac{S(C)-S(A)}{1-\frac{g(q)}{k}S(A)}
\\ \nonumber
&=&\frac{S(C)-S(A)}{1-\frac{g(q)}{k}S(A)}
\end{eqnarray}
or
\begin{eqnarray}                                    \label{d10}
S(C)&=&S(A)+S(B\mid A)-\frac{g(q)}{k}S(A)S(B\mid A),
\end{eqnarray}
which has exactly the same form as Eq.(\ref{d5}). It should be
noticed that this equation means that $S(B\mid A)=S_i(B\mid A)$,
i.e. the conditional entropy of $B$ associated with a microstate
$i$ of $A$ is in fact independent of $i$. It is a little
surprising to see this property with two correlated subsystems.
But if we look back to the origin, we find that this is simply the
consequence of the postulate Eq.(\ref{c6}) which supposes a
nonextensive term in the form of product of two sub-informations.
This is naturally a special choice of the nonextensivity
described by the generalized Hartley formula. If we choose a
different nonextensive information measure, the things will be
different. Recently, the proposition of an entropy
pseudoadditivity\cite{Abe01a} prescribed by thermal equilibrium
shed light on this problem. We understand that, supposed the
generalized Hartley formula (or Tsallis entropy), Eq.(\ref{c6})
and so Eq.(\ref{b12}) are prescribed by the existence of thermal
equilibrium. In other words, without these relations, a composite
system with correlated (not independent!) subsystems can not have
stable equilibrium state. So any exact discussion about
equilibrium systems must conform with the factorization of
compound probability Eq.(\ref{b8}). As a consequence, the idea of
additive energy with neglected correlations\cite{Abe99,Mart01} or
non interacting systems\cite{Guerberoff96}, incompatible with the
spirit of nonextensive statistics, becomes unnecessary and should
be rejected. The definition of temperature has to be revisited on
the basis of nonextensive energy satisfying Eq.(\ref{b8}). It is
what we are doing in the following section.

\section{Thermodynamic relations}

To give statistical interpretation of thermodynamics, a well
defined temperature related to stable thermodynamic equilibrium
(maximum entropy or minimum energy) is needed. In this section, we
will present briefly some consequences of $IS$. All the
discussions are based on Eq.(\ref{b8}) and compatible entropy and
energy pseudoadditivities.

\subsection{Zeroth law and generalized temperature}

First, what is $\beta'$ in the distribution Eq.(\ref{c12})? In
$BGS$, $\beta'=\beta=\frac{1}{kT}$ is the inverse temperature and
the first law of thermodynamics can be written as
\begin{eqnarray}                                    \label{e1a}
dU=dQ+W
\end{eqnarray}
or
\begin{eqnarray}                                    \label{e1}
dU=TdS+YdX
\end{eqnarray}
where W is the work done by $Y$, a generalized exterior force
(e.g. pressure -$P$), $X$ the correspondent displacement (e.g.
volume $V$), and $dS=\frac{dQ}{T}$ the thermodynamic definition of
entropy. When $X$ remains constant, we have $dU=dQ$ and
\begin{eqnarray}                                    \label{e2}
\frac{dS}{dU}=\frac{1}{T}=k\beta.
\end{eqnarray}

On the other hand, within $IS$, from Eqs.(\ref{c7}) and
(\ref{c12}), we obtain
\begin{equation}
S=-k\frac{Z^{-g(q)}-1}{g(q)}+k\beta' Z^{-g(q)}U.    \label{e4}
\end{equation}
or
\begin{equation}
S'=-k\frac{1-Z^{1-q}}{1-q}+k\beta' U.               \label{e4a}
\end{equation}
with $S'=Z^{g(q)}S=Z^{1-q}S$. This leads to
\begin{eqnarray}                                   \label{e5}
\frac{dS'}{dU}=k\beta'
\end{eqnarray}

Now we are showing that $\beta'$ still measures thermal
equilibrium at {\it maximal entropy or minimal energy}. Let us
take the nonextensivity relation Eq.(\ref{d5}) and calculate a
small variance of energy $U(C)$ :
\begin{eqnarray}                                    \label{e6}
\delta U(C) & = & [1-g(q)\beta' U(B)]\delta U(A)+ [1-g(q)\beta'
U(A)]\delta U(B).
\end{eqnarray}
At equilibrium or {\it energy minimum}, $\delta U(C)=0$, we obtain
\begin{eqnarray}                                    \label{e7}
[1-g(q)\beta' U(B)]\delta U(A)=-[1-g(q)\beta' U(A)]\delta U(B).
\end{eqnarray}
Putting this equation into the {\it entropy maximum relation}
Eq.(\ref{b14}) [in which $(1-q)$ should be replaced by $(q-1)$
due to the $IS$ version of entropy nonextensivity Eq.(\ref{d10})],
we get
\begin{eqnarray}                                    \label{e8}
\frac{1-g(q)\beta' U(A)}{1-\frac{g(q)}{k}S(A)}\frac{\partial
S(A)}{\partial U(A)}= \frac{1-g(q)\beta'
U(B)}{1-\frac{g(q)}{k}S(B)}\frac{\partial S(B)}{\partial U(B)}.
\end{eqnarray}
With the help of Eq.(\ref{e4}), we can establish
\begin{eqnarray}                                    \label{e9}
Z^{g(q)}(A)\frac{\partial S(A)}{\partial U(A)}=
Z^{g(q)}(B)\frac{\partial S(B)}{\partial U(B)}
\end{eqnarray}
or
\begin{eqnarray}                                  \label{e9a}
\frac{\partial S'(A)}{\partial U(A)}=\frac{\partial
S'(B)}{\partial U(B)}
\end{eqnarray}
which means
\begin{eqnarray}                                  \label{e10}
\beta'(A)=\beta'(B).
\end{eqnarray}

This result can also be obtained in another way. Multiplying
Eq.(\ref{d10}) by $Z^{1-q}(C)$ and considering $Z(C)=Z(A)Z(B)$,
we obtain
\begin{eqnarray}                                  \label{e10a}
S'(C)&=& Z^{1-q}(B)S'(A)+Z^{1-q}(A)S'(B)-\frac{g(q)}{k}S'(A)S'(B).
\end{eqnarray}
On the other hand, from Eq.(\ref{e4}), it is straightforwardly to
verify that, for maximum entropy at equilibrium, $\delta Z(C)=0$
and $\delta S'(C)=0$. This leads Eq.(\ref{e10a}) to
\begin{eqnarray}                                 \label{e10b}
[Z^{1-q}(B)-\frac{1-q}{k}S'(B)] \delta
S'(A)=-[Z^{1-q}(A)-\frac{1-q}{k}S'(A)] \delta S'(B).
\end{eqnarray}
Now with the help of Eqs.(\ref{e4a}) and (\ref{e7}), we obtain
Eq.(\ref{e9a}) and Eq.(\ref{e10}) .

So $\beta'$ remains the meter of stable equilibrium state. We can
define a generalized temperature
\begin{eqnarray}                                 \label{e11}
T'=\frac{1}{k\beta'}
\end{eqnarray}
where $T'=T$ if $q=1$ or $g(q)=0$.

\subsection{Some other relations}
Considering Eq.(\ref{e5}) and the energy conservation law
Eq.(\ref{e1a}), we see that $dS'$ should be the measure of heat
transfer. We have to write
\begin{eqnarray}                                 \label{e12}
dQ=\frac{dS'}{k\beta'},
\end{eqnarray}
or
\begin{eqnarray}                                 \label{e13}
dQ=T'dS'.
\end{eqnarray}
Now the first law of thermodynamics Eq.(\ref{e1}) should be
written as follows
\begin{eqnarray}                                 \label{e14}
dU=T'dS'+YdX.
\end{eqnarray}
The free-energy $F$ have to be defined as
\begin{eqnarray}                                 \label{e15}
dF=-S'dT'+YdX
\end{eqnarray}
or
\begin{eqnarray}                                 \label{e16}
F=U-T'S',
\end{eqnarray}
which leads to, with the help of Eq.(\ref{e4a}) :
\begin{equation}
F=-kT'\frac{Z^{1-q}-1}{1-q}.                     \label{e17}
\end{equation}
Considering $Z(C)=Z(A)Z(B)$, we easily obtain :
\begin{eqnarray}                                 \label{e17a}
F(C)&=& F(A)+F(B)-\frac{g(q)}{kT'}F(A)F(B).
\end{eqnarray}
We also have, for the heat capacity
\begin{equation}
C_X=\frac{dQ}{dT'}=T' \{ \frac{\partial S'}{\partial T'} \} _X=-T'
\{ \frac{\partial^2 F}{\partial T'^2} \}_X       \label{e18}
\end{equation}
and for the generalized force
\begin{equation}
Y= \{ \frac{\partial F}{\partial X} \}_{T'}.     \label{e19}
\end{equation}

\section{The fundamental problems revisited within incomplete statistics}
What about the problems discussed in section \ref{prob} if we
consider $IS$? Let us examine them one by one.

\begin{enumerate}

\item The first problem of the incompatibility between
normalization and expectation value does not exist any more
because the incomplete normalization Eq.(\ref{c3}) is compatible
with the expectation value Eq.(\ref{c4}).

\item The second problem was that the expectation value
Eq.(\ref{a2a}) was not linear in the distributions shown by
Eq.(\ref{b7}). This problem can be avoided within $IS$ thanks to
Eqs.(\ref{c3}) and (\ref{c4}) :
\begin{eqnarray}                                        \label{f1}
\overline{x}[\lambda p(1)+(1-\lambda)p(2)] &=&\frac{\sum_i[\lambda
p_i^q(1)+(1-\lambda)p_i^q(2)]x_i} {\sum_i[\lambda
p_i^q(1)+(1-\lambda)p_i^q(2)]}  \\   \nonumber
&=&\lambda\sum_ip_i^q(1)x_i+(1-\lambda)\sum_ip_i^q(2)x_i \\
\nonumber &=&\lambda
\overline{x}[p(1)]+(1-\lambda)\overline{x}[p(2)]
\end{eqnarray}

\item The problem of the absence of analytic correlation in energy
and other observable quantities can be resolved by Eqs. (\ref{d4})
and (\ref{d5}), which guarantee a nonextensive statistical theory
with mathematical rigor.

\item The problem of the zeroth law of thermodynamics was resolved
in the above section.

\item The fifth problem discussed in section \ref{prob} is due to
the peculiar equality Eq. (\ref{b20}) which in turn is due to the
average defined in Eq. (\ref{a2a}). In $IS$, this problem does
not exist any more thanks to the incomplete normalization and the
concomitant expectation value Eqs.(\ref{c3}) and (\ref{c4}).

\item In $IS$, the $U-Z$ relation can be found from
Eqs. (\ref{c4}) and (\ref{c17}) :

\begin{eqnarray}                                       \label{f2}
U&=&\sum_{i} p_i^qe_i \\ \nonumber
&=&\frac{1}{Z^q}\sum_{i}e_i[1-(1-q)\beta' e_i]_\dag^\frac{q}{1-q} \\
\nonumber &=&-\frac{1}{Z^q}\frac{\partial}{\partial\beta'}
\sum_{i}[1-(1-q)\beta' e_i]_\dag^\frac{1}{1-q} \\   \nonumber
&=&-\frac{1}{Z^q}\frac{\partial \overline{Z}}{\partial\beta'}
\end{eqnarray}
where
\begin{eqnarray}                                       \label{f3}
\overline{Z}&=& \sum_{i}[1-(1-q)\beta' e_i]_\dag^\frac{1}{1-q}.
\end{eqnarray}

\end{enumerate}

\section{Energy invariance of the distribution}
A crucial problem of the nonextensive distribution Eq.
(\ref{c17}) is that it is not invariant with respect to uniform
translation of energy spectra $e_i$. If we replace $e_i$ by
$e_i+C$ where $C$ is constant, Eq.(\ref{c17}) becomes :

\begin{equation}                                      \label{g1}
p_i=\frac{[1-(1-q)\beta' (e_i+C)]_\dag^\frac{1}{1-q}}{Z}
\end{equation}
with
\begin{equation}                                     \label{g2}
Z^q=\sum_{i}^v[1-(1-q)\beta' (e_i+C)]_\dag^\frac{q}{1-q}
\end{equation}
which is not same as Eq.(\ref{c17}), excepted that $q=1$. This
problem worries enormously scientists\cite{Tsal2}. It is known
that thermostatistics takes into account only energies relative to
thermodynamic movements and that the choice of energy-zero is
never a problem in $BGS$ because the theory has exponential
distribution and is invariant with uniform energy translation.
But with $TDF$, Eq.(\ref{g1}) implies that the properties of a
gas may depend on the translation speed or on the location of the
container. Although in practice we can always choose the usual
energy-zero as for $BGS$, this peculiar theoretical property of
$TDF$ is somewhat unusual and disturbing. Avoiding this puzzling
variance of distribution has been one of the motivations of the
third version of Tsallis theory. Now the following questions
arise : Why do we have to fix only one choice of zero-energy in
$TDF$ to avoid the container-dependence of the statistics? Why
does the nonextensivity lead to this theoretical property? Is it
really something to be avoided?

As is well known, $BGS$ is an extensive theory which holds only
for systems with weak or short range interactions. In addition,
the invariance of $BGS$ is based on the classical mechanics which
leaves the interaction potential completely arbitrary. But $TDF$
is a theory for solving problems with complex interaction or
correlations we often ignore. So we are not obliged or it is not
advisable to identify the variance of $TDF$ to that of $BGS$. The
energy translation invariance is not an universal characteristic
of physical theories. As a matter of fact, this property of
arbitrary potential energy disappears even in classical mechanics
and $BGS$ if we consider the relativistic effect. We can not add
constant into the energy $E$ of a system because $E$ is related
to its total mass $M$ by
\begin{equation}                                     \label{g3}
E=Mc^2
\end{equation}
where $c$ is the light speed. $E$ is not arbitrary because $M$
can not be changed arbitrarily.

If we accept the the variance of $TDF$, we have to choose a
definite zero potential energy. Let us suppose a non negligible
interaction or correlation between all the elements of mass $m_i$
and energy $e_i$ ($i=1, 2, 3, ...$) of a system. Let $V$ be the
total interaction energy. $M$ can be given by :
\begin{equation}                                     \label{g4}
M=\sum_im_i+V/c^2
\end{equation}
or
\begin{equation}                                     \label{g5}
Mc^2=\sum_im_ic^2+V=\sum_ie_i+V
\end{equation}
with $e_i=m_ic^2$ for $i^{th}$ element. It is obvious that neither
$M$, nor $e_i$ and $V$ can be changed. If not, the variance of
the theory would be perturbed \cite{Broglie}. According to this
discussion, a possible choice of zero potential energy corresponds
to the following case
\begin{equation}                                     \label{g6}
M=\sum_im_i
\end{equation}
or $V=0$. This condition may correspond in some cases to infinite
distance between the elements of the system (e.g. an atom when we
consider the internal energy between the electrons and the
nucleus), and in other cases, to special positions of this
elements (e.g. equilibrium position of the atoms in crystal
lattice if we are interested in their vibration). This is in fact
just what we usually choose with $BGS$.

It is worth emphasizing that some of the problems discussed in
section III seem to be related to the distribution functions
invariant through energy translation. Because, if we apply the
maximization method of Martinez et al\cite{Marti}, i.e. to
introduce $\sum_{i}^wp_i^qU$ as the invariance constraint into
the auxiliary function Eq.(\ref{c8}), we obtain :
\begin{equation}                                    \label{c17a}
p_i=\frac{[1-(1-q)\beta' (e_i-U)]_\dag^\frac{1}{1-q}}{Z}
\end{equation}
with
\begin{equation}                                    \label{c18a}
Z^q=\sum_{i}^v[1-(1-q)\beta' (e_i-U)]_\dag^\frac{q}{1-q},
\end{equation}
which is invariant through energy translation and different from
Eq.(\ref{a1}) only by the partition function. We easily find that
the problems 3 to 6 take place again with questionable equalities
similar to Eqs.(\ref{b9}), (\ref{b14}) and (\ref{b20}) which make
it difficult to establish the zeroth law without any
approximation and lead to other puzzling equalities like
Eqs.(\ref{b27}), (\ref{b29a}) and (\ref{b29c4}). Does the
distribution invariance inevitably lead to the theoretical
peculiarities? This seems an interesting topic which is beyond the
range of the present work.

\section{Conclusion}
We have shown some observations about the actual nonextensive
statistical theory. The problems discussed reveal that, with the
conventional normalization Eq.(\ref{a3}) and the expectation
value Eq.(\ref{a2a}), the generalized exponential distribution,
though very successful in many applications, cannot be obtained
with convincing theoretical approach and so the nonextensive
statistics shows peculiar properties which seem difficult to be
avoided. We have shown that it was possible to overcome these
difficulties if we introduced the concept of incomplete
information with suitable normalization and expectation. This
approach allows to establish $TDF$ in a consistent way with only
a different partition function. New nonextensive thermodynamic
relations were deduced on the basis of generalized definitions of
heat and temperature. It is argued that the energy invariance
should not be considered as a necessary property of $TDF$. The
connection between the nonextensivity and the energy shift
dependence of $TDF$ remains to be understood.

\acknowledgments

I acknowledge with great pleasure the very useful discussions with
Professors J.P. Badiali, Shige Peng, Alain Le M\'ehaut\'e and S.
Abe. Thanks are also due to Dr. M. Pezeril, Dr. Laurent Nivanen,
Dr. Fran\c{c}ois Tsobnang, Professor Zengjing Chen and Dr. Yufeng
Shi for valuable discussions.

\end{document}